\documentclass[12pt]{article}

\usepackage{amssymb}
\usepackage[dvips]{graphicx}

\newcommand {\beq}{\begin{equation}}
\newcommand {\eeq}{\end{equation}}
\newcommand {\beqa}{\begin{eqnarray}}
\newcommand {\eeqa}{\end{eqnarray}}
\newcommand {\n}{\nonumber \\}

\newcommand {\Tr}{\mbox{Tr}}

\renewcommand{\theequation}{\thesection.\arabic{equation}}

\begin{document}
\setlength{\oddsidemargin}{0cm}
\setlength{\baselineskip}{7mm}

\begin{titlepage}
\renewcommand{\thefootnote}{\fnsymbol{footnote}}
\begin{normalsize}
\begin{flushright}
\begin{tabular}{l}
OU-HET 535\\
July 2005
\end{tabular}
\end{flushright}
  \end{normalsize}

~~\\

\vspace*{0cm}
    \begin{Large}
       \begin{center}
         {Complex Matrix Model and Fermion Phase Space\\
               for Bubbling AdS Geometries}
       \end{center}
    \end{Large}
\vspace{1cm}

\begin{center}
           Yastoshi T{\sc akayama}\footnote
            {
e-mail address : 
takayama@het.phys.sci.osaka-u.ac.jp}
           {\sc and}
           Asato T{\sc suchiya}\footnote
           {
e-mail address : tsuchiya@phys.sci.osaka-u.ac.jp}\\
      \vspace{1cm}
                    
               {\it Department of Physics, Graduate School of  
                     Science}\\
               {\it Osaka University, Toyonaka, Osaka 560-0043, Japan}
\end{center}

\vspace{3cm}

\begin{abstract}
\noindent
We study a relation between droplet configurations in the bubbling AdS geometries and a complex matrix model that describes the dynamics of a class of chiral primary operators in dual ${\cal N}=4$ super Yang Mills (SYM).
We show rigorously that a singlet holomorphic sector of the complex matrix model is equivalent to a holomorphic part of two-dimensional free fermions,
and establish an exact correspondence between the singlet holomorphic sector of 
the complex matrix model and one-dimensional free fermions.
Based on this correspondence, we find a relation of the singlet holomorphic
operators of the complex matrix model to the Wigner phase space distribution.
By using this relation and the AdS/CFT duality, we give a further
evidence that the droplets in the bubbling AdS geometries
are identified with those in the phase space of the one-dimensional fermions.
We also show that the above correspondence actually maps
the operators
of ${\cal N}=4$ SYM corresponding to the (dual) giant gravitons
to the droplet configurations proposed in the literature.
\end{abstract}
\vfill
\end{titlepage}
\vfil\eject

\setcounter{footnote}{0}

\section{Introduction}
\setcounter{equation}{0}
\renewcommand{\thefootnote}{\arabic{footnote}} 
Recently much attention has been paid to the half-BPS sector in the AdS/CFT
correspondence \cite{Maldacena,GKP,Witten} due to 
the following two interesting observations.
First, it was discussed in \cite{Hashimoto,Jevicki,Berenstein} that
the dynamics of the chiral primary operators of 
${\cal N}=4$ super Yang Mills (SYM) on $S^3\times R$ that correspond to
Kaluza-Klein (KK) gravitons, giant gravitons \cite{giant,BBNS} 
and dual giant gravitons \cite{dual,Hashimoto} in the supergravity side
is described by
a complex matrix model (complex matrix quantum
mechanics), which is obtained by reducing ${\cal N}=4$ SYM to one dimension.
It was pointed out in \cite{Berenstein}
that a holomorphic sector of the
complex matrix model is related to one-dimensional (($1+1$)-dimensional)
fermions in the harmonic potential.
In fact, the one-dimensional free fermions
arise from a hermitian matrix model (hermitian matrix quantum mechanics) 
whose action is formally the same as that of the complex matrix model 
\cite{BIPZ}, and
the number of the degrees of freedom of the hermitian matrix model
is equal to that of the holomorphic sector of the complex matrix model.
Second, as shown by Lin-Lunin-Maldacena (LLM) \cite{LLM}, 
the half-BPS solutions of type IIB supergravity that
preserve $R\times SO(4)\times SO(4)$ isometry 
are characterized by a single function satisfying a differential equation. 
The single function can be determined
by giving an appropriate boundary condition in a two-dimensional subspace 
in ten-dimensional spacetime. 
Giving a boundary condition is equivalent to specifying shapes of droplets 
on the two-dimensional subspace. 
This is so-called  bubbling AdS geometries 
(for further developments, see [12-32]).
The aforementioned gravity states should belong to the above class of the
half-BPS solutions.
It is plausible due to the first observation mentioned above that these
droplets are identified with those in the phase space of the one-dimensional
free fermions.
In fact, some evidences that support this remarkable 
identification have been reported \cite{LLM,Mandal,Grant}.

However, a precise map
between the complex matrix model in the SYM side 
and the droplet configurations in the bubbling
AdS geometries has not been given in the literature.
The purpose of this paper is to find this map and give a further
evidence of the above identification.
Revisiting the analyses in \cite{Jevicki,LLL}, we first show rigorously
that the singlet holomorphic sector of the complex matrix
model is equivalent to a holomorphic part of two-dimensional
($(2+1)$-dimensional) free fermions
in the spherically symmetric harmonic potential.
It has been discussed \cite{IKS1,IKS2} that 
these two-dimensional free fermions are related to  the one-dimensional
free fermions. 
Another interesting work relevant for us is Ref.\cite{Grant}.
The authors of \cite{Grant} identified  fluctuations 
of the droplet of the $AdS_5\times S^5$ geometry that are
responsible for the KK gravitons. Then, by representing the effective action
for the KK gravitons in terms of these fluctuations, they gave a 
supporting evidence of the identification.
Motivated by these works, we next establish an exact correspondence between
the singlet holomorphic sector of the complex matrix model and the 
one-dimensional free fermions. Based on this correspondence, we
find a relation of the singlet holomorphic operators of the complex matrix 
model to the Wigner phase space distribution of the one-dimensional fermions.
Using this relation and the AdS/CFT
duality, we give a further evidence that
the droplets in the bubbling AdS geometries are 
identified with those in
the phase space of the one-dimensional free fermions. 
We also show that the above correspondence actually maps
the operators
of ${\cal N}=4$ SYM corresponding to the (dual) giant gravitons
to the droplet configurations of the supergravity side
proposed in \cite{Berenstein,LLM}. Most of our results also hold 
beyond the classical limit, namely at finite $N$, so that we expect to 
gain a clue from them in extending the bubbling AdS business to finite $N$.

The present paper is organized as follows.
Section 2 is devoted to review. 
In section 2.1, we refer to a class of chiral primary operators of ${\cal N}=4$ SYM on $S^3\times R$ (multi trace operators) and their correlation functions that are analogues of the extremal correlators of ${\cal N}=4$ SYM on $R^4$ and we are concerned with throughout the paper.
We also describe
a complex matrix model that
is used in calculation of these correlation functions.
In section 2.2, based on Refs.\cite{DMW,Mandal,Dhar},
we summarize the results for the one-dimensional 
free fermions in the harmonic potential. 
In particular, we describe the classical phase space density and
its quantum analogue, the Wigner phase space distribution.
In section 2.3, we review the bubbling AdS geometries and its 
conjectured relation
to the classical phase space of the one-dimensional free fermions. 
In section 3, we study the complex matrix model. We reduce the singlet
holomorphic sector of the matrix model to the holomorphic part of 
two-dimensional free fermions in the spherically symmetric harmonic potential.
We also develop the second quantization of these two-dimensional fermions.
This gives an exact correspondence between the singlet holomorphic sector of 
the complex matrix model and the one-dimensional free fermions.
In section 4, based on the correspondence, 
we find a relation of the singlet holomorphic operators of the complex matrix model to the Wigner phase space distribution.
Using this relation and the AdS/CFT duality, we give a further evidence
that the droplets in the bubbling AdS geometries are identified with
those in the phase space of the one-dimensional fermions. 
In section 5, using the correspondence established
in section 3, we construct in the 
Hilbert space of the one-dimensional fermions the states that correspond
to the (dual) giant gravitons, and reproduce
the droplet configurations of the (dual) giant gravitons in
the supergravity side proposed in \cite{Berenstein,LLM}.
Section 6 is devoted to summary and discussion.
In appendix A, we present the results for harmonic oscillators in one and two
dimensions, which we use in the main text.
In appendix B, we give another derivation of the relation in section 4.

\section{Review}
\setcounter{equation}{0}
\subsection{Chiral primary operators and extremal correlators in ${\cal N}=4$ SYM}
In this subsection, we review some facts about a BPS sector of
${\cal N}=4$ SYM. 
We are concerned with the chiral primary operators that take the form
\beqa
{\cal O}^{\{J_1,J_2,\cdots,J_p\}}=
\prod_{a=1}^p \Tr(Z^{J_a}),
\label{chiralprimaryoperators}
\eeqa
where $Z=\frac{1}{\sqrt{2}}(\phi_1+i\phi_2)$ with $\phi_1$ and $\phi_2$
being two of six real scalars of ${\cal N}=4$ SYM. 
Note that these operators are singlet under $U(N)$ transformations and holomorphic, namely depend only on $Z$.
For simplicity, we denote a set $\{J_1,J_2,\cdots,J_p\}$ 
symbolically by $\{J\}$. We first 
consider correlation
functions in ${\cal N}=4$ SYM on $R^4$ of the type
\beqa
\langle ({\cal O}^{\{J^{(0)}\}}(y))^* {\cal O}^{\{J^{(1)}\}}(x_1)
\cdots {\cal O}^{\{J^{(M)}\}}(x_M) \rangle, \;\;\;
y,x_1,\cdots,x_M \in R^4,
\label{extremalcorrelators}
\eeqa
which are called the extremal correlators in the literature.
The charge conservation requires that
\beqa
I^{(0)}=\sum_{r=1}^M I^{(r)},
\eeqa
where 
\beqa
I^{(0)}=\sum_{a_0=1}^{p_0}J_{a_0},\;\;\;
I^{(r)}=\sum_{a_r=1}^{p_r}J_{a_r}.
\eeqa
There is a non-renormalization theorem on the extremal correlators 
telling that one can calculate them 
by using only the free part of the theory \cite{Eden}. Next, we consider
the chiral primary operators (\ref{chiralprimaryoperators}) in
${\cal N}=4$ SYM on $S^3\times R$.
We restrict ourselves to the lowest Kaluza-Klein modes on $S^3$ because
we are interested in half-BPS operators, 
so ${\cal O}^{\{J_1,J_2,\cdots,J_p\}}(t)$ has only the $t$-dependence.
(Linear combinations of) these operators can represent the KK gravitons,
the giant gravitons and the dual giant gravitons 
\cite{Jevicki, Berenstein, LLM}.
We consider correlation functions of the type
\beqa
\langle ({\cal O}^{\{J^{(0)}\}}(t_0))^* {\cal O}^{\{J^{(1)}\}}(t_1)
\cdots {\cal O}^{\{J^{(M)}\}}(t_M) \rangle,
\label{correlationfunctions}
\eeqa
where $t_0 > t_r \;\; (r=1,\cdots,M)$. 
These correlation functions are counterparts of the extremal correlators 
(\ref{extremalcorrelators})
on $R^4$.
It is natural
to consider that the non-renormalization theorem
also holds for the correlation functions
(\ref{correlationfunctions}) and one can calculate them using the free part
of the theory.
Reducing ${\cal N}=4$ SYM to the free part of $Z(t)$
yields a matrix quantum mechanics defined by
\beqa
&&{\cal Z}=\int [dZ(t)dZ^{\dagger}(t)]\: e^{iS}, \n
&&S=\int dt \Tr(\dot{Z}(t)\dot{Z}^{\dagger}(t)-Z(t)Z^{\dagger}(t)),
\label{complexmatrixquantummechanics}
\eeqa
where $Z(t)$ is an $N\times N$ complex matrix and
the path-integral measure is defined through norm 
in the configuration space of the matrix
\beqa
||dZ(t)||^2=2\Tr(dZ(t)dZ^{\dagger}(t)) .
\label{norm}
\eeqa
The potential term in the action arises from a coupling of the conformal
matter to the curvature of $S^3$ and we have rescaled the field and the time
appropriately.
In this paper, we assume the non-renormalization theorem for
the correlation functions (\ref{correlationfunctions}) and concentrate
on their calculation through the complex matrix model 
(\ref{complexmatrixquantummechanics}). The condition 
$t_0 > t_r \;\; (r=1,\cdots,M)$ in (\ref{correlationfunctions}) is
naturally understood from the following fact, 
which was pointed out in \cite{Jevicki}.
The extremal correlators (\ref{extremalcorrelators}) calculated
through the free part of the theory agree with the correlation functions
(\ref{correlationfunctions}) calculated through the matrix model
if all the propagators
$\langle Z_{ij}(x)Z_{kl}^*(y)\rangle
=\delta_{ik}\delta_{jl}/(x-y)^2$ in (\ref{extremalcorrelators}) are replaced
with those of the matrix model 
$\langle Z_{ij}(t)Z_{kl}^*(t_0)\rangle=\delta_{ik}\delta_{jl}e^{i(t-t_0)}/2$
that are valid for $t_0>t$.

\subsection{One-dimensional fermions in the harmonic potential}
In this subsection, based on \cite{DMW,Mandal,Dhar}, 
we summarize the results for a 
one-dimensional system
consisting of $N$ non-interacting fermions in the harmonic
potential. The classical one-body hamiltonian is given by
$h_{cl}=\frac{1}{2}p^2+\frac{1}{2}q^2$.
First, we describe a classical aspect of this system.
A useful object in discussing the classical aspect is
the classical phase space density $u_{cl}(p,q,t)$, which takes the values
$0$ or $1$.
Regions in which $u_{cl}(p,q,t)=1$ are called `droplets'.
The total area of droplets is equal to
$2\pi\hbar N$ (we put $\hbar=1$):
\beqa
\int \frac{dpdq}{2\pi}u_{cl}(p,q,t)=N.
\label{constraintforucl}
\eeqa
It is convenient to introduce the polar coordinates $(r,\phi)$,
\beqa
q=r\cos\phi, \;\;\; p=r\sin\phi.
\eeqa
Then,  $u_{cl}(p,q.t)$ is expressed as 
\beqa
u_{cl}(p,q,t)=\theta(\bar{r}(\phi)-r),
\eeqa
where $\bar{r}(\phi)$ is the boundary profile function, namely the equation
$r=\bar{r}(\phi)$ represents a shape of a boundary of a droplet. 
Here we assume that
there is only a single droplet and $\bar{r}(\phi)$ is a single-valued function.
For example, the ground state is represented by a circular droplet 
$\bar{r}(\phi)=r_0$ with $r_0^2=2N$.
By using the classical equations
of motion $\dot{q}=p,\;\;\dot{p}=-q$, it is easy to show that $u_{cl}$
satisfies the equation of motion
\beqa
\frac{\partial}{\partial t}u_{cl}(p,q,t)
=\frac{\partial}{\partial \phi}u_{cl}(r\cos\phi,r\sin\phi,t)
=\left(q\frac{\partial}{\partial p}-p\frac{\partial}{\partial q}\right)
u_{cl}(p,q,t).
\label{eomforucl}
\eeqa

Next, let us discuss a quantum aspect of the system.
The quantum theory of the system is described by the second-quantized 
hamiltonian 
\beqa
\hat{{\cal H}}=\int dq \psi^{\dagger}(q,t)\left(
-\frac{1}{2}\frac{\partial^2}{\partial q^2}+\frac{1}{2}q^2\right)
\psi(q,t)
\label{secondquantizedhamiltonian}
\eeqa
and a constraint
\beqa
\int dq \psi^{\dagger}(q,t)\psi(q,t)=N,
\label{constraint}
\eeqa
which turns out to fix the total number of particles to $N$.
Here $\psi(q,t)$ is a fermionic field that
satisfies the anti-commutation relation
\beqa
\{\psi(q,t),\psi^{\dagger}(q',t)\}=\delta(q-q').
\label{anti-commutaionrelationbetpsiandpsidagger}
\eeqa
$\psi(q,t)$ obeys the equation of motion
\beqa
i\frac{\partial\psi(q,t)}{\partial t}
=[\psi(q,t),\hat{{\cal H}}]
=\left(-\frac{1}{2}\frac{\partial^2}{\partial q^2}+\frac{1}{2}q^2\right)
\psi(q,t).
\label{eomforpsi}
\eeqa
It is seen from (\ref{eomforpsi}) and 
(\ref{anti-commutaionrelationbetpsiandpsidagger}) 
that $\psi(q,t)$ is expanded in terms
of creation-annihilation operators $\hat{C}_n^{\dagger}$ and $\hat{C}_n$
with an anti-commutation relation 
$\{\hat{C}_m,\hat{C}_n^{\dagger}\}=\delta_{mn}$ as
\beqa
\psi(q,t)=\sum_{n=0}^{\infty}\hat{C}_ne^{-iE_n t}\varphi_n(q),
\label{psi}
\eeqa
where $E_n=n+\frac{1}{2}$ and $\varphi_n(q)$ is the normalized one-body
wave function of the $n$-th excited state, which is defined 
in (\ref{normalizedwavefunction}).
The hamiltonian (\ref{secondquantizedhamiltonian}) and the constraint
(\ref{constraint}) are expressed as
\beqa
&&{\cal H}=\sum_{n=0}^{\infty}E_n \hat{C}_n^{\dagger} \hat{C}_n, \n
&&\sum_{n=0}^{\infty}\hat{C}_n^{\dagger} \hat{C}_n=N.
\eeqa
The left hand side of the second equation is the number operator, as promised.
The ground state is
\beqa
|\Omega\rangle = \hat{C}_{N-1}^{\dagger}\hat{C}_{N-2}^{\dagger}
\cdots \hat{C}_0^{\dagger}|0\rangle,
\eeqa
where $|0\rangle$ is the Fock vacuum defined by $\hat{C}_n|0\rangle=0\;\;
(n=0,1,\cdots)$. Excited states are obtained by replacing
some of the $\hat{C}_n^{\dagger}$ in $|\Omega\rangle$
with $\hat{C}_n^{\dagger}$'s with $n > N-1$ in such a way that the total
number of $\hat{C}_n^{\dagger}$'s is preserved.

An object that plays a crucial role in our analysis is 
the Wigner phase space distribution $\hat{u}(p,q,t)$,
which is defined by
\beqa
\hat{u}(p,q,t)=\int dx e^{ipx} 
\psi^{\dagger}(q+\frac{x}{2},t)\psi(q-\frac{x}{2},t).
\label{uhat}
\eeqa
It follows from (\ref{eomforpsi}) and (\ref{constraint}) that
the Wigner phase space distribution satisfies the equation of motion
\beqa
\frac{\partial}{\partial t}\hat{u}(p,q,t)
=\left(q\frac{\partial}{\partial p}-p\frac{\partial}{\partial q}\right)
\hat{u}(p,q,t)
\label{eomforuhat}
\eeqa
and a constraint
\beqa
\int \frac{dpdq}{2\pi}\hat{u}(p,q,t)=N. 
\label{constraintforuhat}
\eeqa
There is another important constraint on the expectation value of $\hat{u}$,
$\bar{u}(p,q,t)=\langle \Upsilon | \hat{u}(p,q,t) |\Upsilon\rangle$, where
$|\Upsilon\rangle$ is an arbitrary state that satisfies the constraint
(\ref{constraint}). $\bar{u}(p,q,t)$ satisfies a constraint
\beqa
\bar{u}\ast\bar{u}(p,q,t)=\bar{u}(p,q,t),
\label{constraintforubar}
\eeqa
where $*$ stands for the Moyal product:
\beqa
(A \ast B)(p,q)=A(p,q) e^{\frac{i}{2}(
\frac{\stackrel{\leftarrow}{\partial}}{\partial q}
\frac{\stackrel{\rightarrow}{\partial}}{\partial p}
-\frac{\stackrel{\leftarrow}{\partial}}{\partial p}
\frac{\stackrel{\rightarrow}{\partial}}{\partial q})}B(p,q).
\eeqa

It is sometimes convenient to consider a Fourier 
transform of $\hat{u}(p,q,t)$:
\beqa
\tilde{u}(\alpha,\beta,t)=\int \frac{dpdq}{2\pi}e^{-i(p\beta-q\alpha)}
\hat{u}(p,q,t).
\label{utilde}
\eeqa 
It immediately follows from (\ref{eomforuhat}) and (\ref{constraintforuhat})
that $\tilde{u}(\alpha,\beta,t)$
satisfies the equation of motion 
\beqa
\frac{\partial}{\partial t}\tilde{u}(\alpha,\beta,t)
=\left( \beta\frac{\partial}{\partial\alpha}
-\alpha\frac{\partial}{\partial\beta}\right)\tilde{u}(\alpha,\beta,t),
\label{eomforutilde}
\eeqa
and a constraint
\beqa
\tilde{u}(0,0,t)=N.
\label{constraintforutilde}
\eeqa
It is easy to show that $\tilde{u}(\alpha,\beta,t)$ also
satisfies the $W_{\infty}$ algebra
\beqa
&&[\tilde{u}(\alpha,\beta,t),\tilde{u}(\alpha',\beta',t)]
=2i\sin\frac{1}{2}(\alpha\beta'-\alpha'\beta)
\tilde{u}(\alpha+\alpha',\beta+\beta',t).
\label{winfinity}
\eeqa
In fact, $\tilde{u}(\alpha,\beta,t)$ is characterized by
(\ref{eomforutilde}), (\ref{constraintforutilde}) and (\ref{winfinity}).

An important property of $\hat{u}(p,q,t)$ is that its expectation value
$\bar{u}(p,q,t)$ is reduced to
the classical phase space density $u_{cl}(p,q,t)$ in the classical limit.
Actually, because the Moyal product becomes the
ordinary product in the classical limit, 
the constraint (\ref{constraintforubar}) is reduced to 
an equation $(\bar{u})^2=\bar{u}$, whose solution is 
$\bar{u}=0 \;\mbox{or}\;1$.
The equation of motion (\ref{eomforuhat}) and the constraint 
(\ref{constraintforuhat}) also hold for $\bar{u}$, and these equation and 
constraint are equivalent to (\ref{constraintforucl}) and
(\ref{eomforucl}).

\subsection{Half-BPS geometries and free fermion droplets}
In this subsection, we review the LLM description of 
the $AdS_5\times S^5$ geometry and the half-BPS fluctuation modes 
in \cite{LLM,Grant}.

\begin{figure}[htbp]
\begin{center}
 \begin{minipage}{0.45\hsize}
  \begin{center}
   \includegraphics[width=50mm]{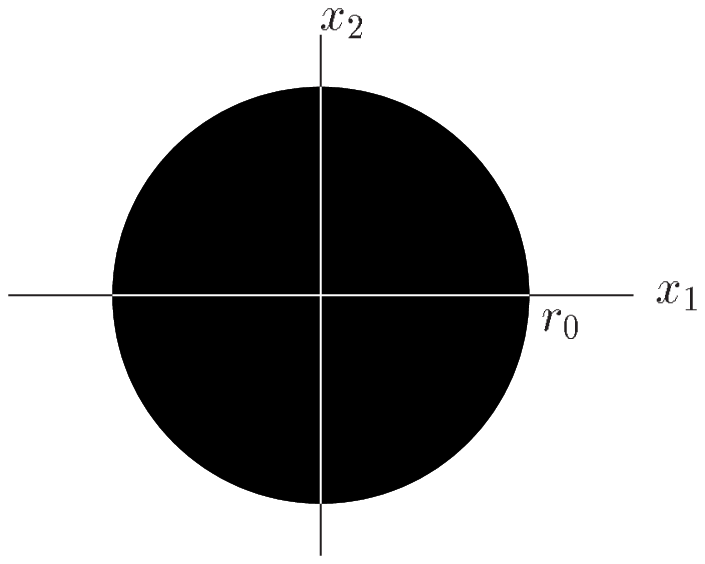}
  \end{center}
  \caption{Circular droplet corresponding to $AdS_5 \times S^5$.}
  \label{fig:circular}
 \end{minipage}
\hspace*{1cm}
 \begin{minipage}{0.45\hsize}
  \begin{center}
   \includegraphics[width=50mm]{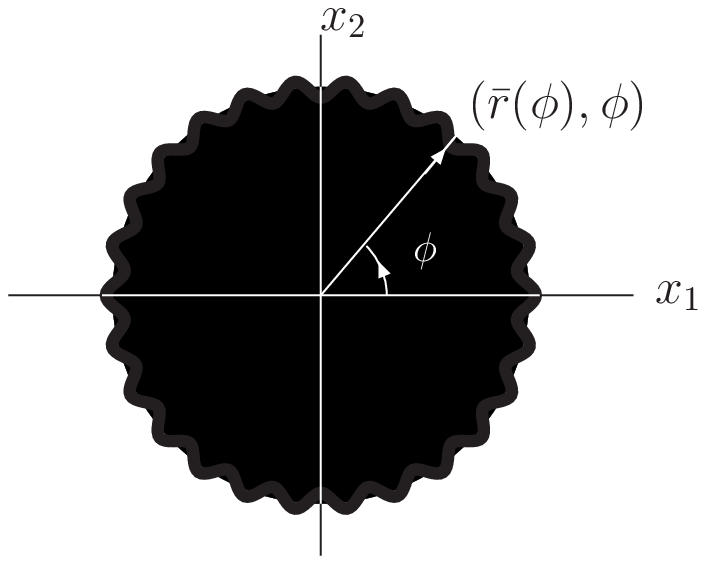}
  \end{center}
  \caption{Fluctuation around the circular droplet.}
  \label{fig:ripple}
 \end{minipage}
\end{center}
\end{figure}

All half-BPS geometries of type IIB supergravity which preserve 
the $R\times SO(4)\times SO(4)$ isometry are obtained 
by LLM \cite{LLM}.
These half-BPS geometries are given by
\begin{eqnarray}
&& ds^2 = -h^{-2}\left[dt + \sum_{i=1}^2V_{i}dx^i\right]^2 
+ h^2 \left[dy^2 + \sum_{i=1}^2dx^idx^i\right] + y e^{G}d\Omega^2_3 + y e^{-G}d\widetilde{\Omega}^2_3\,, \label{LLMmetric} \qquad \\
&& \qquad 
 h^{-2} = 2y\cosh G\,, \quad z= \frac{1}{2}\tanh G\,, 
\nonumber \\ 
&& \qquad y \partial_y V_i = \epsilon_{ij} \partial_j z \,, \quad 
y(\partial_iV_j-\partial_jV_i) = \epsilon_{ij}\partial_yz\,, \label{z-v} \\ 
&& \qquad 
B_t = - \frac{1}{4}y^2 e^{2G}\,, \quad \widetilde{B}_t 
= - \frac{1}{4}y^2 e^{-2G}\,,\nonumber \\ 
&& \qquad F = dB_t\wedge (dt+V)+B_tdV + d\widehat{B}\,, \quad 
 \widetilde{F} = d\widetilde{B}_t\wedge(dt+V) + \widetilde{B}_t dV 
+d\widehat{\widetilde{B}}\,, \nonumber \\
&& \qquad  
d\widehat{B} = -\frac{1}{4}y^3\ast_3d 
\left(\frac{z+\frac{1}{2}}{y^2}\right)\,, 
\quad d\widehat{\widetilde{B}} = - \frac{1}{4}y^3\ast_3 d\left(
\frac{z-\frac{1}{2}}{y^2}\right)\,, \nonumber
\end{eqnarray}
where a single function $z(x^1,x^2,y)$ obeys the differential equation
\begin{eqnarray}
\partial_i\partial_i z + y\partial_y\left(\frac{\partial_y z}{y}\right) = 0\,, 
\label{diff-eq} 
\end{eqnarray} 
and characterizes the solutions.
The function $\frac{1}{2}-z$ on the $y=0$ plane, which we denote by $w$,
takes the values $0$ or $1$ and
defines droplets as $u_{cl}$ does.
In the LLM description, the $AdS_5\times S^5$ geometry is given by
\begin{eqnarray}
{z}(r,y;r_0) 
&=& \frac{r^2-r^2_0+y^2}{2\sqrt{(r^2+r^2_0+y^2)^2-4r^2r^2_0}}\,,
\\\nonumber
V_r &=&0\,,
\qquad
V_{\phi} = -\frac{1}{2}
\left[ \frac{r^2 + r_0^2 + y^2}{\sqrt{(r^2+r^2_0+y^2)^2-4r^2r^2_0}}-1\right]\,.
\end{eqnarray}
Here $r_0$ is identified with the AdS radius $R$ 
through $r_0 = R^2=\sqrt{2N}$\, and $(r,\phi)$ are the polar coordinates of the $x_1$-$x_2$ plane.
The droplet configuration ($w(x_1,x_2)$) 
is a circular droplet with the radius $r_0$ in the $(x_1,x_2)$ plane (see
Fig. \ref{fig:circular}).

In \cite{Grant} a subclass of metric perturbation around 
the $AdS_5\times S^5$ is discussed.
Such perturbation is obtained by considering 
small ripples of the circular droplet.
The boundary of the perturbed droplet in the polar coordinates is given by
\begin{eqnarray}
\bar{r}(\phi)=r_0+\delta{r(\phi)}.
\label{rbar}
\end{eqnarray}
$\delta{r(\phi)}$ is expanded in Fourier modes as
\begin{eqnarray}
\delta{r(\phi)}=\sum_{n \neq 0}a_n e^{in\phi},
\quad
a_n^*=a_{-n}
\label{deltar}
\end{eqnarray}
(see Fig. \ref{fig:ripple}).
Here the zero mode is absent due to the area preserving deformation 
(to the first order in $\delta{r}$).
Then one can express the resultant metric perturbation in terms of $a_n$.
The result shows that 
the gravity modes described by $a_n$ are the KK gravitons that,
according to the standard AdS/CFT dictionary, correspond to
$\Tr Z^n$ for $n >0$ and $\Tr(Z^{\dagger|n|})$ for $n<0$ in dual ${\cal N}=4$
SYM. Moreover, one can read off the commutation relation
among $a_n$'s from the effective action that gives
the equations of motion for this metric perturbation: 
\beqa
[a_m,a_n] \sim m\delta_{m+n}.
\eeqa
This commutation relation agrees with the one that is obtained when
the droplet is regarded as the droplet on the classical phase space of 
one-dimensional fermions in 
the harmonic potential, with the identification
$x_1=q,\;\;x_2=p$. Thus we are allowed to identify droplets in the $x_1-x_2$
plane with those in the $(p,q)$ phase space of the one-dimensional fermions.


\section{Complex matrix model}
\setcounter{equation}{0}
\subsection{Complex matrix model and free fermions in 
two dimensions}
In this subsection, we study the complex matrix model introduced
in section 2.2. In particular, we show that if we restrict ourselves to 
the correlation functions (\ref{correlationfunctions}), we can regard
the model as a holomorphic part of a two-dimensional system of 
free fermions in the spherically symmetric harmonic potential. 
The action in (\ref{complexmatrixquantummechanics}) 
is evaluated as
\beqa
S=\int dt \Tr(\dot{Z}(t)\dot{Z}^{\dagger}(t)-Z(t)Z^{\dagger}(t))
=\int dt \sum_{i,j}(\dot{Z}(t)_{ij}\dot{Z}(t)_{ij}^*
-Z(t)_{ij}(t)Z(t)_{ij}^*),
\label{cqmqmaction}
\eeqa
while the norm (\ref{norm}) as
\beqa
||dZ(t)||^2=
2\Tr(dZ(t)dZ^{\dagger}(t))=2\sum_{i,j}dZ(t)_{ij}dZ(t)^*_{ij},
\eeqa
from which we find an explicit form of the path-integral measure:
\beqa
[dZ(t)dZ^{\dagger}(t)]
=[2^{N^2}\prod_{i,j}d\mbox{Re}Z(t)_{ij}d\mbox{Im}Z(t)_{ij}].
\label{path-integralmeasure}
\eeqa
By comparing (\ref{cqmqmaction}) and (\ref{path-integralmeasure}) with
(\ref{partitionfunctionfor2dharmonicoscillator}) 
and (\ref{pathintegralmeasurefor2dharmonicoscillator}) 
in appendix A, we see that the system is 
nothing but a set of $N^2$ independent two-dimensional harmonic oscillators
that are spherically symmetric. 

We can quantize the system
canonically as we do for the one-body system in appendix A.
In what follows, we make use of the results in appendix A.
The quantum hamiltonian is
\beqa
\hat{H}=\sum_{i,j}\left(-\frac{\partial^2}{\partial Z_{ij}\partial Z_{ij}^*}
+Z_{ij}Z_{ij}^* \right),
\label{Hhat}
\eeqa
and the normalized ground state wave function is
\beqa
\chi_0=\frac{1}{\pi^{\frac{N^2}{2}}}e^{-\Tr(ZZ^{\dagger})}
=\frac{1}{\pi^{\frac{N^2}{2}}}e^{-\sum_{ij}Z_{ij}Z_{ij}^*}.
\eeqa
The measure for the inner product between the wave functions is
\beqa
\int \prod_{ij}dZ_{ij}dZ_{ij}^*
=2^{N^2}\int \prod_{ij}d\mbox{Re}Z_{ij}d\mbox{Im}Z_{ij}.
\eeqa

We see from the results in appendix A that wave functions that take the form
\beqa
\chi^{(J_1,\cdots,J_K)}=\left(\prod_{b=1}^{K}\Tr(Z^{J_b})\right)\chi_0
\label{holomorphicwavefn}
\eeqa
are eigenfunctions of $\hat{H}$ with eigenvalues $N^2+\sum_{b=1}^K J_b$.
This observation enables us to calculate the
correlation functions (\ref{correlationfunctions}) as follows:
\beqa
&&\langle ({\cal O}^{\{J^{(0)}\}}(t_0))^* {\cal O}^{\{J^{(1)}\}}(t_1)
\cdots {\cal O}^{\{J^{(M)}\}}(t_M) \rangle \n
&&=\langle\chi_0 |({\cal O}^{\{J^{(0)}\}}(t_0))^* 
{\cal O}^{\{J^{(1)}\}}(t_1)\cdots {\cal O}^{\{J^{(M)}\}}(t_M)|\chi_0\rangle \n
&&=\langle\chi_0 | e^{i\hat{H}t_0}({\cal O}^{\{J^{(0)}\}})^*
e^{-i\hat{H}(t_0-t_1)}{\cal O}^{\{J^{(1)}\}}e^{-i\hat{H}(t_1-t_2)}
\cdots e^{-i\hat{H}(t_{M-1}-t_M)}{\cal O}^{\{J^{(M)}\}}e^{-i\hat{H}t_M}
|\chi_0\rangle \n
&&=e^{i\sum_{r=1}^M I^{(r)} (t_r-t_0)}
\int \prod_{ij}dZ_{ij}dZ_{ij}^* 
({\cal O}^{\{J^{(0)}\}}\chi_0)^* ({\cal O}^{\{J^{(1)}\}}
\cdots{\cal O}^{\{J^{(M)}\}}\chi_0).
\label{calculationofcorrelationfunctions}
\eeqa
Here the result is independent of the order of $t_1,\cdots,t_M$ although
we assumed in the above calculation $t_0>t_1>\cdots >t_M$.
Hence it is necessary and sufficient for calculating the correlation functions
(\ref{correlationfunctions})
to examine the wave functions (\ref{holomorphicwavefn}), which are
holomorphic except for the factor $e^{-\sum_{i,j}Z_{ij}Z_{ij}^*}$ and singlet under $U(N)$ transformations.

Let us see that as far as
the wave functions (\ref{holomorphicwavefn}) are concerned, we can
reduce the dynamical degrees of freedom of the system to 
$N$ eigenvalues of $Z$.
We express $Z$ in terms of a unitary matrix and a triangle complex 
matrix as 
\beqa
Z=UTU^{\dagger},
\label{diagonalization}
\eeqa
where $UU^{\dagger}=1$, $T_{ij}=0$ for $i>j$ and the $T_{ii}\;(i=1,\cdots,N)$
are the eigenvalues of $Z$ \cite{Mehta}. 
Note that the number of real parameters in $Z$ is $2N^2$ 
while those in $U$ and $T$ are $N^2$ and $N(N+1)$, respectively.
This reflects the fact that
$Z$ does not determine $U$ and $T$ uniquely: if $V$ is a unitary 
diagonal matrix, $(UV)^{\dagger}Z(UV)=V^{\dagger}TV$ and
$V^{\dagger}TV$ is also a triangle 
matrix. This redundancy enables us to impose $N$ constraints on
variation of $U$. Variation of $Z$ is
\beqa
dZ=U(dT+i(dHT-TdH))U^{\dagger},
\eeqa
where $dH=-iU^{\dagger}dU$ and $dH$ is a hermitian matrix. 
We impose the constraint $dH_{ii}=0\; (i=1,\cdots,N)$.
Because $||dZ||^2=||dT+i(dHT-TdH)||^2$, we can make a change of variables from
$Z_{ij}$ to $H_{ij} \; (i>j)$ and $T_{ij}\; (i\leq j)$. The jacobian for
this change of variables is $|\Delta(z)|^2$ \cite{Mehta},
where
\beqa
\Delta(z)=\prod_{i<j}(z_i-z_j),\;\;\;\; z_i\equiv T_{ii}\;\;\;(i=1,\cdots,N).
\eeqa
Hence we obtain
\beqa
\int \prod_{i,j}dZ_{ij}dZ_{ij}^*
=\int \prod_{i>j}dH_{ij}dH_{ij}^* \prod_{k<l}dT_{kl}dT_{kl}^*
\prod_m dz_m dz_m^* |\Delta(z)|^2.
\eeqa
This leads us to redefine a wave function $\chi$ as 
\beqa
\chi_F\equiv \Delta(z)\chi
\eeqa
and the hamiltonian $\hat{H}$ as
\beqa
\hat{H}_F\equiv \Delta(z)\hat{H}\frac{1}{\Delta(z)}
\eeqa
Thus the system is equivalently 
described by the hamiltonian $\hat{H}_F$ and the wave
functions $\chi_F$ with the inner product
\beqa
\langle \chi^{(1)}_F | \chi^{(2)}_F\rangle
=\int\prod_{i>j}dH_{ij}dH_{ij}^*\prod_{k<l}dT_{kl}dT_{kl}^*
\prod_m dz_mdz_m^* \chi_F^{(1)*}\chi_F^{(2)}.
\label{innerproductbetchiandchi}
\eeqa
In particular, if $\chi$ is an eigenstate of $\hat{H}$ with an eigenvalue $E$,
$\chi_F$ is an eigenstate of $\hat{H}_F$ with the same eigenvalue $E$.
$\chi^{(J_1,\cdots,J_K)}_F$ is
expressed in terms of $z_i$ and $T_{ij} \;(i<j)$ as
\beqa
&&\chi^{(J_1,\cdots,J_K)}_F=\Delta(z)\chi^{\left( J_1,\cdots,J_K \right)}
=\Delta(z)\left(\prod_{a=1}^K \sum_{i_a}z_{i_a}^{J_a}\right)\chi_0,
\label{chiF}\\
&&\chi_0=\frac{1}{\pi^{\frac{N^2}{2}}}
e^{-\sum_i z_iz_i^*-\sum_{j<k}T_{jk}T_{jk}^*}. \nonumber
\eeqa
It should be noted that this eigenstate is independent of $H_{ij}$ and $H_{ij}^*$ due to the singlet nature of the wave function (\ref{holomorphicwavefn}), while this eigenstate is independent of $T_{ij}$, $T_{ij}^*$, and $z_i^*$ except for $\chi_0$ due to the holomorphy and the singlet nature of the wave function (\ref{holomorphicwavefn}).
We see from this expression that
$\chi^{(J_1,\cdots,J_K)}_F$ is equal to a certain linear combination of
\beqa
&&\left|\begin{array}{cccc}
\Phi_{l_1}(z_1,z_1^*)&\Phi_{l_1}(z_2,z_2^*)&\cdots&\Phi_{l_1}(z_N,z_N^*)\\
\Phi_{l_2}(z_1,z_1^*)&                     &      &\vdots               \\
\vdots               &                     &      &\vdots               \\
\Phi_{l_N}(z_1,z_1^*)&\cdots               &\cdots&\Phi_{l_N}(z_N,z_N^*)
\end{array}\right| 
\times\prod_{j<k}\Phi_0(T_{jk},T_{jk}^*) \n
&&\mbox{with}\;\; \sum_i l_i=\frac{1}{2}N(N-1)+\sum_{b=1}^K J_b,
\label{productofwavefn}
\eeqa
where $\Phi_l(z,z^*)$ is a `holomorphic' wave function or a wave function of
the lowest Landau level, which is defined in (\ref{Phil}).
Namely, $\chi^{(J_1,\cdots,J_K)}_F$ is a wave function of an energy
eigenstate of a two-dimensional system consisting of 
$N$ fermions and $\frac{1}{2}N(N-1)$ bosons
which are governed by the spherically symmetric
harmonic potential and do not interact each other.
In this picture the coordinates of the fermions are represented by $(z_i,z_i^*)$, while the coordinates of the bosons are represented by $(T_{jk}, T_{ji}^*)$.
Here, the fermions are in the ``holomorphic" states, and 
the bosons are in the ground state. The total energy of the fermions is 
$\frac{1}{2}N(N+1)+\sum_{b=1}^KJ_b$, where $\frac{1}{2}N(N+1)$ corresponds to
the energy of the ground state of fermions, 
and the total energy of the boson is $\frac{1}{2}N(N-1)$.
The sum of these two total energies is equal to
the eigenvalue of $\hat{H}_F$ with respect to $\chi^{(J_1,\cdots,J_K)}_F$,
$N^2+\sum_{b=1}^KJ_b$, so that we can identify the system with the system
of these free fermions and bosons, as far as we concentrate on 
the wave functions $\chi^{(J_1,\cdots,J_K)}_F$.
Because the bosons are always in the ground state,
their contribution to the eigenvalues of $\hat{H}_F$ is always
$\frac{1}{2}N(N-1)$, and in the inner product
$\langle \chi^{(J^{(1)}_1,\cdots,J^{(1)}_{K_1})}_F | 
\chi^{(J^{(2)}_1,\cdots,J^{(2)}_{K_2})}_F\rangle$
the integral over $H_{ij}$ and $T_{ij}$ gives an overall constant factor.
Therefore we are allowed to consider only the fermionic part of 
$\chi^{(J_1,\cdots,J_K)}_F$ and replace the measure in the inner product
(\ref{innerproductbetchiandchi}) with $\int \prod_idz_idz_i^*$.
In this way, we have reduced the singlet holomorphic sector of
the complex matrix model to the holomorphic 
part of two-dimensional free fermions in the harmonic potential. 
This holomorphic nature enables us to relate these two-dimensional fermions to the one-dimensional ones, as we will see in the next subsection.

Finally we make a comment on the singlet non-holomorphic sector which is non-BPS.
The hamiltonian (\ref{Hhat}) describes a set of $N^2$ non-interacting harmonic oscillators represented by $(Z_{ij}, Z_{ij}^*)$.
After change of variables from $(Z_{ij},Z_{ij}^*)$ to $(z_i, z_i^*, T_{ij}, T_{ij}^*, H_{ij}, H_{ij}^*)$, the hamiltonian is described by the dynamical degrees of freedom represented by $(z_i, z_i^*, T_{ij}, T_{ij}^*, H_{ij}, H_{ij}^*)$ that are complicatedly interacting.
Nevertheless, as we showed above, the singlet holomorphic sector is reduced to the non-interacting fermions and bosons represented by $(z_i, z_i^*)$ and $(T_{ij}, T_{ij}^*)$ respectively.
If we consider the singlet non-holomorphic sector, such simplification does not occur.
We need to treat the above hamiltonian with the complicated interactions.

\subsection{Second quantization of two-dimensional fermions}
In this subsection, we develop a second quantized theory of the holomorphic
part of the two-dimensional free fermions in the previous subsection and
establish an exact correspondence between the singlet holomorphic sector of the 
complex matrix model and the one-dimensional free fermions.
The observation in the previous subsection leads us to introduce a
fermion field
\beqa
\Psi(z,z^*,t)=\sum_{l=0}^{\infty}\hat{C}_l 
e^{-iE_lt}\Phi_l(z,z^*),
\label{Psi}
\eeqa
where $E_l=l+1$, and
$\hat{C}_l$ and $\hat{C}_l^{\dagger}$ satisfy the anti-commutation relation
$\{\hat{C}_l,\hat{C}_m^{\dagger}\}=\delta_{lm}$. The $\hat{C}_l$ in
(\ref{Psi}) will be identified with $\hat{C}_l$ in (\ref{psi}) below. 
This fermion field satisfies
an anti-commutation relation
\beqa
\{\Psi(z,z^*,t),\Psi^{\dagger}(z',{z'}^*,t)\}
=\sum_{l=0}^{\infty}\Phi_l(z,z^*)\Phi_l(z',{z'}^*)
\equiv K(z,z^*;z',{z'}^*).
\eeqa
$K$ behaves as the delta function with respect to functions spanned by
the $\Phi_l(z,z^*)$, namely holomorphic functions of $z$ times $e^{-zz^*}$.
$\Psi(z,z^*,t)$ also satisfies a constraint that the total number of fermions
be $N$,
\beqa
\int dzdz^* \Psi^{\dagger}(z,z^*,t)\Psi(z,z^*,t)
=\sum_{l=0}^{\infty}\hat{C}^{\dagger}_l\hat{C}_l
=N.
\eeqa
The ground state $|\Omega\rangle$ is again given by
\beqa
|\Omega\rangle=\hat{C}_{N-1}^{\dagger}\hat{C}_{N-2}^{\dagger}\cdots
\hat{C}_1^{\dagger}\hat{C}_0^{\dagger} |0\rangle,
\eeqa
where $|0\rangle$ is the Fock vacuum defined by $\hat{C}_l|0\rangle=0$.

We define the operators $\hat{s}_J(t)$ by
\beqa
\hat{s}_J(t)=\int dzdz^* z^J \hat{U}(z,z^*,t),\;\;\;\;
\hat{U}(z,z^*,t)=\Psi^{\dagger}(z,z^*,t)\Psi(z,z^*,t),
\label{sJhat}
\eeqa
where $\hat{U}(z,z^*,t)$ is interpreted as the density operator restricted
to the ``holomorphic" sector.
It is easy to verify that one can calculate the correlation functions 
(\ref{correlationfunctions})
by making a correspondence in the second line of 
(\ref{calculationofcorrelationfunctions}):
\beqa
\Tr(Z(t)^J)           \; &\leftrightarrow& \; \hat{s}_J(t) \n
\Tr(Z^{\dagger}(t)^J) \; &\leftrightarrow& \; \hat{s}_J^{\dagger}(t) \n
|\chi_0\rangle        \; &\leftrightarrow& \; |\Omega\rangle.
\label{correspondence}
\eeqa
The explicit form of $\hat{s}_J(t)$ is easily calculated as
\beqa
\hat{s}_J(t)=\frac{1}{2^{\frac{J}{2}}}e^{iJt}\sum_{l=0}^{\infty}
\sqrt{\frac{(l+J)!}{l!}}\hat{C}_{l+J}^{\dagger}\hat{C}_l.
\label{explicitsJhat}
\eeqa
Note that because $[\hat{s}_J(t),\hat{s}_{J'}(t)]=0$, there is no ordering
ambiguity in translating the multi-trace operators into
products of $\hat{s}_J$'s. 
Thus, if we identify $\hat{C}_l$ and $\hat{C}^{\dagger}_l$ in $\hat{s}_J$
with those of the one-dimensional fermions,
we establish an exact correspondence between the singlet holomorphic sector of 
the complex matrix model and the one-dimensional free fermions in the
harmonic potential.

\begin{figure}[htbp]
\begin{center}
 \begin{minipage}{0.45\hsize}
  \begin{center}
   \includegraphics[width=35mm]{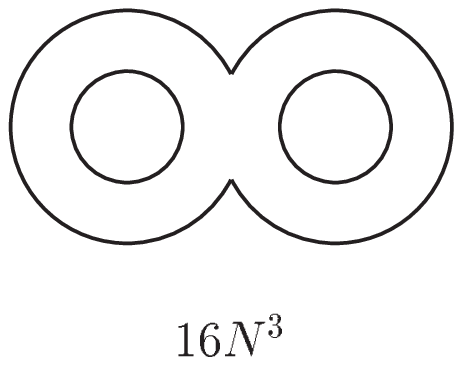}
  \end{center}
  \caption{planar contribution.}
  \label{fig:planar}
 \end{minipage}
\hspace*{1cm}
 \begin{minipage}{0.45\hsize}
  \begin{center}
   \includegraphics[width=35mm]{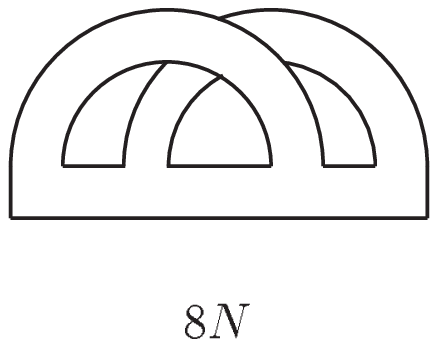}
  \end{center}
  \caption{nonplanar contribution.}
  \label{fig:nonplanar}
 \end{minipage}
\end{center}
\end{figure}

As an example, we give the calculation of 
$\langle\Tr(Z^{\dagger}(t_0)^4)\Tr(Z(t_1)^2)\Tr(Z(t_2)^2)\rangle$ with
$t_0>t_1,\;t_2$, based on the formalism we developed above.
\beqa
\langle\Tr(Z^{\dagger}(t_0)^4)\Tr(Z(t_1)^2)\Tr(Z(t_2)^2)\rangle 
&=&\langle\Omega | \hat{s}^{\dagger}_4(t_0)\hat{s}_2(t_1)
\hat{s}_2(t_2)|\Omega\rangle \n
&=&\frac{1}{2^4}e^{2i(t_1-t_0)+2i(t_2-t_0)}
\langle\Omega | PQR |\Omega\rangle,
\label{example}
\eeqa
where
\beqa
P&=&\sqrt{(N+3)(N+2)(N+1)N}\hat{C}^{\dagger}_{N-1}\hat{C}_{N+3}
+\sqrt{(N+2)(N+1)N(N-1)}\hat{C}^{\dagger}_{N-2}\hat{C}_{N+2} \n
&&+\sqrt{(N+1)N(N-1)(N-2)}\hat{C}^{\dagger}_{N-3}\hat{C}_{N+1} 
  +\sqrt{N(N-1)(N-2)(N-3)}\hat{C}^{\dagger}_{N-4}\hat{C}_N\n
Q&=&\sqrt{(N-1)(N-2)}\hat{C}^{\dagger}_{N-1}\hat{C}_{N-3}
+\sqrt{(N-2)(N-3)}\hat{C}^{\dagger}_{N-2}\hat{C}_{N-4} \n
&&+\sqrt{(N+3)(N+2)}\hat{C}^{\dagger}_{N+3}\hat{C}_{N+1}
+\sqrt{(N+2)(N+1)}\hat{C}^{\dagger}_{N+2}\hat{C}_{N}, \n
R&=&\sqrt{(N+1)N}\hat{C}^{\dagger}_{N+1}\hat{C}_{N-1}
+\sqrt{N(N-1)}\hat{C}^{\dagger}_{N}\hat{C}_{N-2}.
\eeqa
$\mbox{}$From (\ref{example}), we obtain the final result
\beqa
\frac{16N^3+8N}{2^4}e^{2i(t_1-t_0)+2i(t_2-t_0)},
\label{finalresult}
\eeqa
which we can also obtain by applying the `Wick theorem' with respect to
$Z_{ij}$ and $Z_{ij}^*$ to the last line of 
(\ref{calculationofcorrelationfunctions}). Note that the final result
contains the non-planar contribution 
(See Figs. \ref{fig:planar},\ref{fig:nonplanar}).
\footnote{(\ref{finalresult}) is valid
for $U(N)$ gauge group. For $SU(N)$, the subleading term
is modified \cite{deMelloKoch}.}

\section{Singlet holomorphic operators \\ \hfill and the Wigner phase space distribution}
\setcounter{equation}{0}
In this section, based on the exact correspondence between the singlet holomorphic
sector of the complex matrix model and the one-dimensional fermions,
which was established in the previous section,
we find an exact relation of the singlet holomorphic operators of the complex matrix model to the Wigner phase space distribution of the one-dimensional fermions.
The relation gives via the AdS/CFT duality an evidence that
the droplets in the bubbling AdS geometries are identified
with those in the phase space of 
the one-dimensional free fermions in the harmonic potential.

In section 2.3, we saw $a_n\;(n>0)$ in (\ref{deltar}),
which is the Fourier mode of the small fluctuation around the circular droplet
(Fig \ref{fig:ripple}),
yields the KK graviton that corresponds to $\Tr(Z(t)^n)$ in dual
${\cal N}=4$ SYM according to the standard AdS/CFT dictionary.
In this case, $w$ in section 2.3 that specifies the droplet
is given by
\beqa
w(x_1,x_2)&=&\theta(r_0+\delta r(\phi)-r) \n
&=&\theta(r_0-r)+\delta (r_0-r)\delta r(\phi)+{\cal O}((\delta r)^2),
\eeqa
where $x_1=r\cos\phi,\;\;x_2=r\sin\phi$. Thus, we obtain from (\ref{deltar})
a relation
\beqa
a_n=\frac{1}{2\pi r_0g_n(r_0)}\int dr \int d\phi r g_n(r) e^{-in\phi}
w(x_1,x_2).
\label{an}
\eeqa
At this stage, $g_n(r)$ is arbitrary. 
In the previous section,
we constructed $\hat{s}_J(t)$ which
corresponds to $\Tr(Z(t)^J)$ quantum mechanically. 
The quantum analogue of the classical phase space density 
$u_{cl}$ is $\hat{u}$.
Then, if the droplet specified by $w(x_1,x_2)$ are identified with that
in the phase space of the free fermions through
$w(x_1,x_2)=u_{cl}(p,q)$ with
$x_1=q$ and $x_2=p$, we expect the following identity to hold quantum 
mechanically: 
\beqa
\hat{s}_J(t)=A_J\int dr \int d\phi r g_J(r) e^{-iJ\phi} \hat{u}(p,q,t).
\label{sJhatbyuhat}
\eeqa
Here an overall coefficient $A_J$ and a function $g_J(r)$ will be determined
soon. The existence of the relation (\ref{sJhatbyuhat}) serves
as an evidence that the droplets in the bubbling AdS
geometries are identified with those in the phase space of the one-dimensional
free fermions. This is one of our results in this paper.

In what follows, we will see that the relation (\ref{sJhatbyuhat}) indeed
holds.
We first evaluate $\hat{u}(p,q,t)$ explicitly. From (\ref{psi}) 
and (\ref{uhat}), we obtain
\beqa
\hat{u}(p,q,t)
=\sum_{m,n=0}^{\infty}\hat{C}_m^{\dagger}\hat{C}_n e^{i(m-n)t}
\int dx e^{ipx} \varphi_m^*(q+\frac{x}{2})\varphi_n(q-\frac{x}{2}).
\label{calculationofuhat}
\eeqa
In order to calculate the integral in (\ref{calculationofuhat}),
we deform the contour as $x\rightarrow 2x+2ip$ and 
make use of the following formula:
\beqa
\int dx e^{-x^2}H_m(x+v)H_n(x+w)
=2^n\pi^{\frac{1}{2}}m!w^{n-m}L_m^{n-m}(-2vw) \;\;\;\mbox{for} \;\;m\leq n,
\eeqa
where $L_{n}^{n-m}$ is the Laguerre polynomial defined by
\beqa
L_n^{\alpha}(x)=\sum_{m=0}^n (-1)^m 
\left(\begin{array}{c}n+\alpha\\n-m\end{array}\right)
\frac{x^m}{m!}.
\eeqa
The final result is
\beqa
\hat{u}(p,q,t)
&=&\sum_{m\leq n}\hat{C}^{\dagger}_m\hat{C}_n e^{i(m-n)t}
\sqrt{\frac{m!}{n!}}2^{\frac{n-m}{2}+1}(-1)^m e^{i(m-n)\phi} e^{-r^2}
r^{n-m}L_m^{n-m}(2r^2) \n
&&+\sum_{m>n}\hat{C}^{\dagger}_m\hat{C}_n e^{i(m-n)t}
\sqrt{\frac{n!}{m!}}2^{\frac{m-n}{2}+1}(-1)^n e^{i(m-n)\phi} e^{-r^2}
r^{m-n}L_n^{m-n}(2r^2).
\label{explicituhat}
\eeqa 
By comparing (\ref{explicitsJhat}) with (\ref{sJhatbyuhat}) and
(\ref{explicituhat}), we find that (\ref{sJhatbyuhat}) holds if
$A_J=\frac{1}{2\pi 2^J}$ and $g_J(r)=r^J$. Namely, we find
\beqa
\hat{s}_J(t) &=& \int dzdz^* z^J \hat{U}(z,z^*,t) \n
&=&\frac{1}{2\pi 2^J}\int dr \int d\phi r^{J+1} e^{-iJ\phi} \hat{u}(p,q,t) \n
&=&\int \frac{dpdq}{2\pi}\left(\frac{q-ip}{2}\right)^J\hat{u}(p,q,t).
\label{sJhatanduhat}
\eeqa
$\mbox{}$From (\ref{sJhatanduhat}),
we naively expect that a relation 
$\hat{U}(z,z^*,t)=\frac{1}{\pi}\hat{u}(p,q,t)$
holds with the identification $z=\frac{1}{2}(q-ip)$. Unfortunately, this
is not the case. It is, however, known \cite{IKS2} that there is a simple
relation between $\tilde{u}(\alpha,\beta,t)$ and a Fourier transform of 
$\hat{U}$. In appendix B, we give another derivation of 
(\ref{sJhatanduhat}) based on this relation.


\section{Mapping (dual) giant gravitons to droplets}
\setcounter{equation}{0}
In this section, using the correspondence between the singlet holomorphic sector
of the complex matrix model and the one-dimensional free fermions, which
was established in section 3, we construct in the 
Hilbert space of the one-dimensional fermions the states corresponding to
the (dual) giant gravitons.
We will see that the classical phase space
density that is the classical limit of the expectation value
of the Wigner phase space distribution with respect to these states
actually agree with the corresponding 
droplet configurations in the supergravity side proposed 
in \cite{Berenstein,LLM}.

We first consider a sequence of non-negative integers
$\langle\lambda\rangle=(\lambda_1,\lambda_2,\cdots,\lambda_N)$ with
$\lambda_1 \geq \lambda_2 \geq \cdots \geq \lambda_N \geq 0$
and $\sum_{i=1}^N \lambda_i=n$. This sequence defines a Young tableau
with the number of boxes in the $i$-th row being $\lambda_i$ and the total
number of boxes being $n$, and 
specifies an irreducible representation 
of $GL(N,C)$ as well as an irreducible representation of
the symmetry group $S_n$.
As in \cite{Jevicki}, we introduce an operator which is
a linear combination of the chiral primary operators 
(\ref{chiralprimaryoperators}):
\beqa
\chi_{\langle\lambda\rangle}(Z)
=\frac{1}{n!}\sum_{\sigma \in S_n}\chi_{\langle\lambda\rangle}(\sigma)
\mbox{tr}(\sigma Z),
\label{chilambdaZ}
\eeqa
where $\chi_{\langle\lambda\rangle}(Z)$ and 
$\chi_{\langle\lambda\rangle}(\sigma)$ are characters
of the representations specified by $\langle\lambda\rangle$, and
$\mbox{tr}(\sigma Z)$ is defined by
\beqa
\mbox{tr}(\sigma Z)=\sum_{i_1,i_2,\cdots,i_n=1}^N 
Z_{i_1i_{\sigma(1)}}Z_{i_2i_{\sigma(2)}}\cdots Z_{i_ni_{\sigma(n)}}.
\eeqa
As in \cite{Jevicki}, we make use of Weyl's character formula.
Weyl's character formula tells that (\ref{chilambdaZ}) is equal to
the Schur polynomial:
\beqa
\chi_{\langle\lambda\rangle}(Z)
=\frac{\det_{i,j}z_j^{N-i+\lambda_i}}{\Delta(z)}.
\eeqa
Hence a fermionic wave function that is obtained by acting the operator
(\ref{chilambdaZ}) on the ground state is given by
\beqa
\chi_F \sim
\left|\begin{array}{cccc}
\Phi_{N-1+\lambda_1}(z_1,z_1^*)&\Phi_{N-1+\lambda_1}(z_2,z_2^*)
&\cdots&\Phi_{N-1+\lambda_1}(z_N,z_N^*)\\
\Phi_{N-2+\lambda_2}(z_1,z_1^*)&          &      &\vdots               \\
\vdots               &                     &      &\vdots               \\
\Phi_{\lambda_N}(z_1,z_1^*)&\cdots     &\cdots&\Phi_{\lambda_N}(z_N,z_N^*)
\end{array}\right| 
\times\prod_{j<k}\Phi_0(T_{jk},T_{jk}^*). 
\eeqa
The correspondence established in section 3 maps this wave function to
a state in the Hilbert space of the one-dimensional free fermions
\beqa
\hat{C}^{\dagger}_{N-1+\lambda_1}\hat{C}^{\dagger}_{N-2+\lambda_2}
\cdots \hat{C}^{\dagger}_{1+\lambda_{N-1}}\hat{C}^{\dagger}_{\lambda_N}
|0\rangle.
\label{state}
\eeqa
This state can also be obtained by rewriting (\ref{chilambdaZ}) in terms
of $\hat{s}_J$ through (\ref{correspondence}) and acting the resultant
operator on the ground state $|\Omega\rangle$.

\begin{figure}[htbp]
\begin{center}
 \begin{minipage}{0.45\hsize}
  \begin{center}
   \includegraphics[width=50mm]{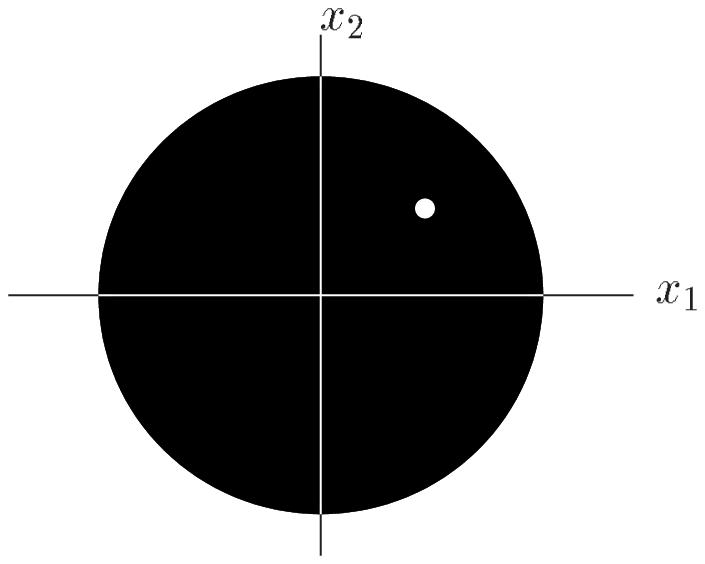}
  \end{center}
  \caption{A giant graviton in $AdS_5 \times S^5$.}
  \label{fig:giant}
 \end{minipage}
\hspace*{1cm}
 \begin{minipage}{0.45\hsize}
  \begin{center}
   \includegraphics[width=50mm]{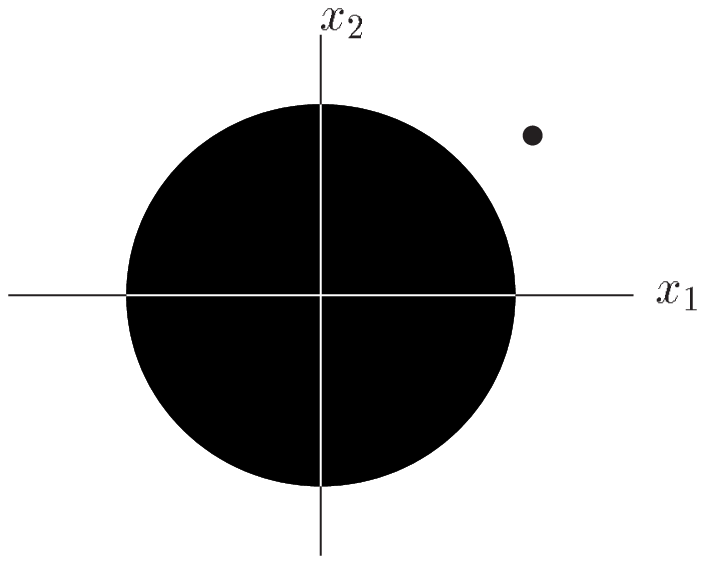}
  \end{center}
  \caption{A dual giant graviton in $AdS_5 \times S^5$.}
  \label{fig:dualgiant}
 \end{minipage}
\end{center}
\end{figure}

Next, let us construct in the Hilbert space of the one-dimensional fermions
the states that represent the (dual) giant gravitons.
In ${\cal N}=4$ SYM side, 
the giant graviton with the angular momentum $L$, where $L$ is compatible
with $N$, corresponds to
the operator (\ref{chilambdaZ}) 
with $\langle\lambda\rangle=(\lambda_1=1,\lambda_2=1,\cdots,\lambda_L=1,
\lambda_{L+1}=0,\cdots,\lambda_N=0)$ \cite{BBNS,Jevicki}. 
This $\langle\lambda\rangle$ corresponds
to the Young tableau that consists of a single column with
$L$ boxes. 
Note that there is a restriction $L \leq N$ and the giant graviton with $L=N$
is called the maximal giant graviton.
We see from (\ref{state}) that
the state that is obtained by acting this operator
on the ground state is mapped to a state in the Hilbert space of the 
one-dimensional fermions
\beqa
|GG;L\rangle = \hat{C}^{\dagger}_N\hat{C}_{N-L}|\Omega\rangle.
\label{giantgraviton}
\eeqa
The phase space density that is the classical limit of
$\langle GG;L|\hat{u}(p,q,t)|GG;L\rangle$
obviously specifies
the circular droplet with a small circular defect inside the circle
(see Fig. \ref{fig:giant}). This droplet configuration is indeed
proposed in \cite{Berenstein,LLM}
as that of the giant graviton in the bubbling AdS geometries.

The dual giant graviton with angular momentum $L$, where $L$ is
compatible with or much larger than $N$, corresponds to
the operator (\ref{chilambdaZ}) with 
$\langle\lambda\rangle=(L,0,\cdots,0)$ \cite{Jevicki}. 
This $\langle\lambda\rangle$
corresponds to the Young tableau that consists of a single row with $L$ boxes.
We again see from (\ref{state}) that the state that is obtained by acting 
this operator on the ground state is mapped to a state in the Hilbert space
of the one-dimensional fermions
\beqa
|DGG;L\rangle=\hat{C}^{\dagger}_{N-1+L}\hat{C}_{N-1}|\Omega\rangle.
\label{dualgiantgraviton}
\eeqa
This is also consistent with the proposal in \cite{Berenstein,LLM} that
the dual giant gravitons are given by the circular droplet and a small 
droplet outside the circular droplet in the bubbling AdS geometries
(see Fig. \ref{fig:dualgiant}).

\section{Summary and discussion}
\setcounter{equation}{0}
In this paper, we study a relation between the droplet configurations
in the bubbling AdS geometries and the singlet holomorphic sector of the complex matrix
model that can describe a class of the chiral primary operators 
of ${\cal N}=4$ SYM.
We first showed rigorously that the singlet holomorphic sector of the complex matrix
model is equivalent to the holomorphic part of the two-dimensional fermions.
We developed the second quantization of these fermions and established the
exact correspondence between the singlet holomorphic sector of the complex matrix
model and the one-dimensional free fermions in the harmonic potential.
Next, based on this correspondence, we find a relation between 
the operators $\Tr(Z(t)^J)$ of the complex matrix model 
and the Wigner phase space distribution.
This gives
via the AdS/CFT duality
a further evidence that the droplets in the bubbling AdS geometries
are identified with those in the phase space of the one-dimensional free
fermions. Finally, we also constructed the states corresponding to the
(dual) giant gravitons in the Hilbert space of the one-dimensional fermions.
We obtained the droplet configurations for these states, which are consistent
with the proposals in \cite{Berenstein,LLM}.
Our main results are summarized in (\ref{correspondence}), 
(\ref{explicitsJhat}), (\ref{sJhatanduhat}), (\ref{giantgraviton}) 
and (\ref{dualgiantgraviton}).
These relations also hold at finite $N$, so that we expect to gain a 
clue from them in extending the bubbling AdS business to finite $N$.


We make a comment on a hermitian matrix
model whose action is formally the same as the complex matrix model:
\beqa
{\cal Z}=\int [d\phi(t)] e^{iS}, \;\;\;
S=\int dt \left(\frac{1}{2}\dot{\phi}(t)^2-\frac{1}{2}\phi(t)^2\right),
\eeqa
where the path-integral measure is defined by the norm 
$||d\phi(t)||^2=\Tr((d\phi(t))^2)$.
As is well-known \cite{BIPZ}, as far as the $U(N)$-invariant sector 
is concerned,
the model can be reduced to a system of $N$ eigenvalues and these eigenvalues
behaves as one-dimensional non-interacting fermions in the harmonic potential,
which were discussed in section 2.2. 
Then, a $U(N)$-invariant operator 
$\Tr(\phi(t)^J)$ which is a counterpart of $\Tr(Z(t)^J)$ in the complex 
matrix model is translated to a operator in the second quantized theory
of the fermions
\beqa
\hat{v}_J(t)=\int dq q^J \psi^{\dagger}(q,t)\psi(q,t).
\eeqa
We cannot, however, replace $\hat{s}_J$ with $\hat{v}_J$ in the calculation
of the correlation functions (\ref{correlationfunctions}). This yields extra
terms that are formally obtained by `contracting' $Z_{ij}$ and $Z_{kl}$.
This is a manifestation of the difference between the complex and hermitian
matrix models.

There are some directions as extension of the present work.
First, it has been discussed that
the pp-wave limit of the bubbling AdS space is described by 
relativistic fermions in $(1+1)$ dimensions \cite{LLM,Takayama}. 
It is important to find
a relation between these fermions and the complex matrix model. 
Topology changes
in the bubbling geometry are also an interesting subject \cite{Horava}. 
We hope our findings
in this paper
to give some insight to this subject. 

\section*{Acknowledgements}
We would like to thank Y. Susaki, T. Takayanagi and K. Yoshida for discussions.
The work of Y.T. is supported in part by The 21st Century COE Program
``Towards a New Basic Science; Depth and Synthesis."
The work of A.T. is supported in part by Grant-in-Aid for Scientific
Research (No.16740144) from the Ministry of Education, Culture, Sports,
Science and Technology.

\section*{Appendix A: Harmonic oscillators in one and two dimensions}
\setcounter{equation}{0}
\renewcommand{\theequation}{A.\arabic{equation}}
In this appendix, we summarize the results for the harmonic oscillators
in one and two dimensions, which we use throughout this paper.
The harmonic oscillator in one dimension is defined by a hamiltonian,
\beqa
\hat{h}=-\frac{1}{2}\frac{\partial^2}{\partial q^2}+\frac{1}{2}q^2
=\hat{a}^{\dagger}\hat{a}+\frac{1}{2},
\eeqa
where $\hat{a}^{\dagger}$ and $\hat{a}$ are of course the creation and
annihilation operators:
\beqa
\hat{a}=\frac{1}{\sqrt{2}}(q+\frac{\partial}{\partial q}),\;\;\;
\hat{a}^{\dagger}=\frac{1}{\sqrt{2}}(q-\frac{\partial}{\partial q}),
\eeqa
which satisfy the commutation relation $[\hat{a},\hat{a}^{\dagger}]=1$.
The ground state wave function $\varphi_0(q)$ is characterized by a relation
$\hat{a}\phi_0(q)=0$ and the normalized one takes the form
\beqa
\varphi_0(q)=\frac{1}{\pi^{\frac{1}{4}}}e^{-\frac{1}{2}q^2}.
\eeqa
The normalized wave function of the $n$-th excited state, whose energy 
eigenvalue is $n+\frac{1}{2}$, is explicitly given by
\beqa
\varphi_{n}(q)=\frac{1}{\sqrt{n!}}(\hat{a}^{\dagger})^n \varphi_0(q)
=\frac{1}{\sqrt{\pi^{\frac{1}{2}}2^n n!}}H_n(q)e^{-\frac{1}{2}q^2},
\label{normalizedwavefunction}
\eeqa
where $H_n$ is the Hermite polynomial.

We generalize the above results to the
two-dimensional spherically symmetric harmonic oscillator, which is defined
by a hamiltonian,
\beqa
\hat{h}=-\frac{1}{2}\frac{\partial^2}{\partial q_1^2}
-\frac{1}{2}\frac{\partial^2}{\partial q_2^2}
+\frac{1}{2}q_1^2+\frac{1}{2}q_2^2
=\hat{a}_1^{\dagger}\hat{a}_1+\hat{a}_2^{\dagger}\hat{a}_2+1,
\eeqa
where
\beqa
\hat{a}_i=\frac{1}{\sqrt{2}}(q_i+\frac{\partial}{\partial q_i}),\;\;\;
\hat{a}_i^{\dagger}=\frac{1}{\sqrt{2}}(q_i-\frac{\partial}{\partial q_i}),
\;\;\;(i=1,2).
\eeqa
These creation and annihilation operators satisfy the commutation relations
\beqa
[\hat{a}_i,\hat{a}_j^{\dagger}]=\delta_{ij},\;\;\;
[\hat{a}_i,\hat{a}_j]=[\hat{a}_i^{\dagger},\hat{a}_j^{\dagger}]=0.
\eeqa
The normalized wave function for the $(m,n)$ excited state, 
whose energy eigenvalue is $m+n+1$, is given by 
$\varphi_{m,n}(q_1,q_2)=\varphi_m(q_1)\varphi_n(q_2)$.

It is crucial for our analysis in the main text to introduce the complex
coordinate:
\beqa
z=\frac{1}{{\sqrt{2}}}(q_1+i q_2).
\eeqa
In this variable, the hamiltonian takes the form
\beqa
\hat{h}=-\frac{\partial^2}{\partial z \partial z^*}+zz^*
=\hat{c}_1^{\dagger}\hat{c}_1+\hat{c}_2^{\dagger}\hat{c}_2+1,
\eeqa
where
\beqa
&&\hat{c}_1=\frac{1}{\sqrt{2}}(\hat{a}_1+i\hat{a}_2)
=\frac{1}{\sqrt{2}}(z+\frac{\partial}{\partial z^*}),\;\;\;
\hat{c}_1^{\dagger}
=\frac{1}{\sqrt{2}}(\hat{a}_1^{\dagger}-i\hat{a}_2^{\dagger})
=\frac{1}{\sqrt{2}}(z^*-\frac{\partial}{\partial z}),\n
&&\hat{c}_2=\frac{1}{\sqrt{2}}(\hat{a}_1-i\hat{a}_2)
=\frac{1}{\sqrt{2}}(z^*+\frac{\partial}{\partial z}),\;\;\;
\hat{c}_2^{\dagger}
=\frac{1}{\sqrt{2}}(\hat{a}_1^{\dagger}+i\hat{a}_2^{\dagger})
=\frac{1}{\sqrt{2}}(z-\frac{\partial}{\partial z^*}),
\eeqa
which satisfy again the commutation relations
$[\hat{c}_i,\hat{c}^{\dagger}_j]=\delta_{ij}, \;\;\;
[\hat{c}_i,\hat{c}_j]=[\hat{c}_i^{\dagger},\hat{c}_j^{\dagger}]=0$.
By acting the creation operators $\hat{c}_1^{\dagger}$ and 
$\hat{c}_2^{\dagger}$ successively on the ground state, one can construct 
energy eigenstates, which form a normalized orthonormal basis:
\beqa
\Phi_{k,l}(z,z^*)
=\frac{1}{\sqrt{k!l!}}(\hat{c}_1^{\dagger})^k(\hat{c}_2^{\dagger})^l
\varphi_{0,0}.
\eeqa
The ground state, whose energy eigenvalue is one, takes the form in the 
complex variable
\beqa
\Phi_{0,0}(z,z^*)=\frac{1}{\sqrt{\pi}}e^{-zz^*}.
\eeqa
The excited state $\Phi_{k,l}(z,z^*)$ with the energy eigenvalue $k+l+1$
is expressed as a linear combination of $\varphi_{m,n}$ with 
$m+n=k+l$. In particular, $\Phi_{0,l}$ is holomorphic except for the factor
$e^{-zz^*}$: 
\beqa
\Phi_{0,l}(z,z^*)=\sqrt{\frac{2^l}{\pi l!}}z^l e^{-zz^*}.
\label{Phil}
\eeqa
We denote $\Phi_{0,l}(z,z^*)$ simply by $\Phi_{l}(z,z^*)$. 
$\Phi_{l}(z,z^*)$ is a wave function of the lowest Landau level in the 
context of the quantum Hall effect.
The measure used
for the inner product between the wave functions is
\beqa
\int dq_1 dq_2=2\int d\mbox{Re}z d\mbox{Im}z^* \equiv \int dz dz^*.
\eeqa
The system can be quantized equivalently by path-integral. The partition
function is given by
\beqa
Z=\int[dz(t)dz^*(t)]e^{-S},\;\;\;\;
S=\int dt (\dot{z}(t)\dot{z}^*(t)-z(t)z^*(t)),
\label{partitionfunctionfor2dharmonicoscillator}
\eeqa
where 
\beqa
[dz(t)dz^*(t)]=[2d\mbox{Re}z(t)d\mbox{Im}z(t)].
\label{pathintegralmeasurefor2dharmonicoscillator}
\eeqa

\section*{Appendix B: Another derivation of (\ref{sJhatanduhat})}
\setcounter{equation}{0}
\renewcommand{\theequation}{B.\arabic{equation}}
In this appendix, we give another derivation of (\ref{sJhatanduhat}).
We first consider a Fourier transform
of $\hat{U}(z,z^*,t)$
\beqa
\tilde{U}(\Lambda,\Lambda^*,t)
=\int dzdz^* e^{\Lambda^* z-\Lambda z^*} \hat{U}(z,z^*,t), 
\eeqa
which is an analogue of $\tilde{u}$. 
By substituting (\ref{Psi}), we obtain
\beqa
\tilde{U}(\Lambda,\Lambda^*,t)
=\sum_{m,n=0}^{\infty}\hat{C}_m^{\dagger}\hat{C}_n e^{i(m-n)t}
\sqrt{\frac{2^{m+n}}{m!n!}}(-1)^m
\frac{\partial^{m+n}}{\partial \Lambda^m \partial \Lambda^{*n}}
e^{-\frac{1}{2}\Lambda\Lambda^*}.
\eeqa
Some algebra gives
\beqa
&&[\tilde{U}(\Lambda,\Lambda^*,t),\tilde{U}(\Lambda',{\Lambda'}^*,t)] \n
&&=(e^{\frac{1}{2}\Lambda^*\Lambda'}-e^{\frac{1}{2}\Lambda{\Lambda'}^*})
\tilde{U}(\Lambda+\Lambda',\Lambda^*+{\Lambda'}^*,t) \n
&&=e^{\frac{1}{4}(\Lambda+\Lambda')(\Lambda^*+{\Lambda'}^*)
-\frac{1}{4}\Lambda\Lambda^*-\frac{1}{4}\Lambda'{\Lambda'}^*}
2i\sin\frac{1}{4}(\Lambda^*\Lambda'-\Lambda{\Lambda'}^*)
\tilde{U}(\Lambda+\Lambda',\Lambda^*+{\Lambda'}^*,t)
\label{algebra}
\eeqa
This result urges us to define a new object 
\beqa
\tilde{U}_{new}(\Lambda,\Lambda^*,t)
=e^{\frac{1}{4}\Lambda\Lambda^*}\tilde{U}(\Lambda,\Lambda^*,t).
\eeqa
In fact, $\tilde{U}_{new}(\Lambda,\Lambda^*,t)$ satisfies 
the $W_{\infty}$ algebra (\ref{winfinity}) with
the identification $\Lambda=\beta-i\alpha$. It is also easy to see that
under this identification
$\tilde{U}_{new}(\Lambda,\Lambda^*,t)$ satisfies 
the same equation of motion (\ref{eomforutilde}) 
and constraint (\ref{constraintforutilde}) 
as $\tilde{u}(\alpha,\beta,t)$ satisfies. 
Therefore $\tilde{U}_{new}$ would coincide with $\tilde{u}$: 
\beqa
\tilde{u}(\alpha,\beta,t)=e^{\frac{1}{4}\Lambda\Lambda^*}
\tilde{U}(\Lambda,\Lambda^*,t). 
\label{utildeandUhat}
\eeqa
We also checked this by explicit calculation. 
The algebra (\ref{algebra}) and the relation (\ref{utildeandUhat}) were
already derived in \cite{IKS2} in the context of the integer quantum Hall 
effect. Thus we obtain a direct relation
between $\hat{U}$ and $\hat{u}$:
\beqa
\hat{U}(z,z^*,t)=\int \frac{d\Lambda d\Lambda^*}{4\pi^2}
e^{-\Lambda^*z+\Lambda z^*-\frac{1}{4}\Lambda\Lambda^*}
\int \frac{dpdq}{2\pi}e^{-\frac{\Lambda}{2}(q+ip)+\frac{\Lambda^*}{2}(q-ip)}
\hat{u}(p,q,t).
\label{Uhatanduhat}
\eeqa
Using (\ref{Uhatanduhat}), we calculate $\hat{s}_J(t)$ as
\beqa
\hat{s}_J(t)&=&\int dzdz^* z^J \hat{U}(z,z^*,t) \n
&=&\int dzdz^*\int \frac{d\Lambda d\Lambda^*}{4\pi^2}(-1)^J
\frac{\partial^J}{\partial \Lambda^{*J}}(e^{-\Lambda^*z+\Lambda z^*})
e^{-\frac{1}{4}\Lambda\Lambda^*}
\int \frac{dpdq}{2\pi}e^{-\frac{\Lambda}{2}(q+ip)+\frac{\Lambda^*}{2}(q-ip)}
\hat{u}(p,q,t) \n
&=&\left.\frac{\partial^J}{\partial \Lambda^{*J}}\left(
e^{-\frac{1}{4}\Lambda\Lambda^*}
\int \frac{dpdq}{2\pi}e^{-\frac{\Lambda}{2}(q+ip)+\frac{\Lambda^*}{2}(q-ip)}
\hat{u}(p,q,t) \right)\right|_{\Lambda=\Lambda^*=0}.
\eeqa
The last expression actually gives (\ref{sJhatanduhat}).


\end{document}